\documentclass[conference]{IEEEtran}

\usepackage{graphicx}
\usepackage[pdftex,colorlinks=true,pdfstartview=FitV,linkcolor=black,citecolor=black,urlcolor=black,bookmarks=false]{hyperref}
\usepackage{xcolor}

\begin{document}
%
\title{LikeStarter: a Smart-contract based\\ Social DAO for Crowdfunding\footnotemark}

\author{
\IEEEauthorblockN{Mirko Zichichi, Michele Contu, Stefano Ferretti, Gabriele D'Angelo}
\IEEEauthorblockA{
Department of Computer Science and Engineering\\
University of Bologna\\
Bologna, Italy\\
\href{mailto:mirko.zichichi@studio.unibo.it}{mirko.zichichi@studio.unibo.it},
\href{mailto:michele.contu@studio.unibo.it}{michele.contu@studio.unibo.it},
\href{mailto:s.ferretti@unibo.it}{s.ferretti@unibo.it},
\href{mailto:g.dangelo@unibo.it}{g.dangelo@unibo.it}
}
}

\maketitle

\footnotetext{The publisher version of this paper is available at \url{https://doi.org/10.1109/INFCOMW.2019.8845133}.
\textbf{{\color{red}Please cite this paper as: ``Mirko Zichichi, Michele Contu, Stefano Ferretti, Gabriele D'Angelo. LikeStarter: a Smart-contract based Social DAO for Crowdfunding. Proceedings of the 2st Workshop on Cryptocurrencies and Blockchains for Distributed Systems (CryBlock'19)''.}}}

\begin{abstract}
Crowdfunding has become a popular form of collective funding, in which small donations or investments, made by groups of people, support the development of new projects in exchange of free products or different types of recognition. 
Social network sites, on the other hand, promote user cooperation and currently are at the basis of any individuals cyber-interactions. 
In this paper, we present LikeStarter, a blockchain-based decentralized platform that combines social interactions with crowdfunding mechanisms, allowing any user to raise funds while becoming popular in the social network. 
Being built over the Ethereum blockchain, LikeStarter is structured as a Decentralized Autonomous Organization (DAO), that fosters crowdfunding without the intervention of any central authority, and recognizes the active role of donors, enabling them to support artists or projects, while making profits.
\end{abstract}

%
\IEEEpeerreviewmaketitle

%
%

%

\section{Introduction}
Social media platforms are universally recognized as an important medium for conveying and spreading information all around the world. By their nature, it is trivial to reach out a broad audience and at the same time engage users by increasing their connectedness and fostering their direct participation in the communication process~\cite{KIM2018153}.
In this perspective, the union between social interaction and crowdfunding represents a powerful symbiosis. In fact, the social impact on the public obtained by social interactions can be used as a way to raise funds. Many examples exist of unknown people or companies that get funded by other people, based on their social interactions, thanks to services such as Twitch, Patreon, IndieGoGo, Kickstarter. These ``creators'' of contents are supported by funders, who are entertained (or convinced) by their creations and motivated to participate to the donation, in order to seek rewards and enforce the connections with people in their social network~\cite{gerber}.
Moreover, social networks such as Steemit witness the value of the social users' activities. In this like-economy, users' likes and sharing add more value to any kind of memes, data and social profiles.

On the other hand, blockchain technologies have changed radically the vision that we have of the Internet. More specifically, the blockchain has drastically evolved the way that we used to consider the finance, trust in communication and even renewed the concept of digital democracy~\cite{buterin2013}. Many of the concepts in the vision of hyper-connect and trusted world were already emerging as common ideas, but the blockchain has offered the tools for their fast realization, without requiring the presence of a third party. Thus, this technology allowed the fast growth of decentralized currencies, self-executing digital contracts (i.e.~smart contracts) and intelligent assets that can be controlled over the Internet (i.e.~smart property), reinforced also by the social impact it has had in society.
The blockchain also enables the development of new governance systems characterized by more democratic and inclusive decision-making. Decentralized Autonomous Organizations (DAOs), once deployed in the blockchain, live and operate without the need to any human intermediary~\cite{article}.

In this paper, we present LikeStarter, a social network site where users can raise funds for other users through a simple ``like'', built on top of the Ethereum blockchain. LikeStarter allows an individual that has something to propose to be crowdfunded easily through the use of smart contracts. Here, ``something'' implies that every kind of project or self-proposal can be promoted by a user, through the use of the LikeStarter site. 
LikeStarter is a DAO, that employs smart contracts to control and manage funds.

The rest of this paper is organized as follows. Section \ref{sec:back} provides the background. Section \ref{sec:like} presents the LikeStarter DAO, while in \ref{sec:archi} we provide a description of its architecture. Section \ref{sec:use} presents a use case with artists crowdfunding, and shows some features of the related Decentralized Application (DAPP). Finally, Section \ref{sec:conc} provides some concluding remarks.

\section{Background}\label{sec:back}

This section presents some of the main concepts needed to introduce the LikeStarter application.
 
 \subsection{Ethereum} 
 

The blockchain is a decentralized implementation of a distributed ledger that records a set of transactions generated by multiple users. Besides being a novel version of a distributed, pseudo-anonymized and untamperable database, the presence of smart contracts make such system a novel distributed, replicated computing model, with a certain complexity \cite{gda-jpdc-2017}.

Ethereum is a popular blockchain that promotes the development of decentralized applications, based on a Turing-complete scripting language~\cite{buterin2013}. Ethereum consists of a shared virtual machine that belongs to anyone, but that is governed by no one. It is permanently available, no one can interrupt the execution or to censor it \cite{D'Angelo:2018}. The execution of public code is carried on by multiple nodes that are part of the network. 

Ethereum has a related cryptocurrency named Ether. Ether can be transferred between accounts through transactions and it is used to compensate miner nodes who maintain the blockchain evolution. However, Ethereum is not only a cryptocurrency; in fact, Ethereum introduces four main features in its system:
\begin{itemize}
    \item Tokens: different currencies living in the blockchain;
    \item Smart Contracts: digital contracts where rules are stated through code;
    \item Smart Property: a way to assert the ownership of a real (non-digital) asset;
    \item DAO: a Decentralized Autonomous Organization, structured as a set of smart contracts that define tokens, properties and government regulations of the organization.
\end{itemize}

 \subsection{Initial Coin Offerings (ICOs)}
 
 Initial Coin Offerings (ICOs) are public offers of new cryptocurrencies, that can be obtained in exchange for existing ones~\cite{8327568}. In an ICO, some amount of a cryptocurrency is transformed into tokens, which are functional to financing projects or activities in the blockchain ecosystem. Token systems have many applications, e.g.~sub-currencies representing assets (such as USD, gold, etc.), company stocks, tokens representing shares for a smart property, secure unforgeable coupon~ \cite{buterin2013}.

In Ethereum, the realization of a standard token typically follows the ERC-20 interface~\cite{erc20}, a guideline that specified how standard tokens should be organized. More specifically, ERC-20 defines the functions and events that an Ethereum token contract has to implement. These guidelines include the way in which tokens are transferred between addresses and the methods through which the data contained within them is accessible.
 
 \subsection{Smart Contracts}

Smart Contracts are a fundamental component of the Ethereum framework. They are computer programs residing on the blockchain, which are triggered through specific transactions. They are not governed by a central authority. Moreover they are trackable and cannot be modified, once they are deployed on the blockchain. 
Smart contracts are business logics being semi-autonomously executed to move currency and force payments as a result of an agreement. Contracts can store currency, data and executable code. Moreover, smart contracts can communicate with other contracts and even create new ones. Contracts are reactive entities, i.e.~they are triggered by external events coming from external accounts (i.e.~human users)~\cite{Seijas2016ScriptingSC}.

 \subsection{Decentralized Autonomous Organization (DAO)}
  A Decentralized Autonomous Organization (DAO) is a virtual entity managed by smart contracts and executed in a decentralized way. The use of the Ethereum blockchain implies that the organization state is maintained by a consensus system and that contracts are used to implement transactions, currency flows, rules and rights inside of the organization.
  
After a DAO code is deployed, members can interact through smart contracts and Ether may be sent or received.
In exchange for Ether, a DAO usually uses tokens that are assigned to the sender account. Tokens grant its holder different rights and, moreover, their quantity is proportional to the amount of Ether that has been transferred.

Members of a DAO are able to propose options for decision in the organization but they can also discuss and vote those through transparent mechanisms. Within the contracts, individual actions of members cannot be directly determined~\cite{jentzsch2016}.
 
 \subsection{Crowdfunding Mechanisms}
 
 Crowdfunding, and in particular equity crowdfunding, is a mechanism that enables broad groups of investors to fund startup companies or individuals in return for equity. The crowdfunding has been used for a wide range of for-profit, entrepreneurial ventures such as artistic and creative projects, medical expenses, travel, or community-oriented social entrepreneurship projects~\cite{GLEASURE2016101}.
 
 Unlike the traditional way of fundraising, where a company receive a small number of major investments from a restricted group of lenders, in crowdfunding, a company or an individual receive support through a large number of small contributions from multiple customers. As a consequence, crowdfunding represents a mixture of entrepreneurship and social network participation, where customers play also the role of investors~\cite{chunta}.
Investors funding liability increases proportionally with accumulated capital and may lead to ``herding effect''~\cite{NBERw19133}, in which users are more encouraged to donate if the campaign is successful.
In fact, social media mechanics are also used to identify promising projects that are more likely to be successful so as to create more profits from the initial investment.
 
 The landmarks of our work and successful examples of crowdfunding platforms are Kickstarter, regarding the crowdfunding of startup companies and, more importantly, Patreon, which shows how the influence of an individual in a social network can lead to a prosperous crowdfunding.

 \section{LikeStarter}\label{sec:like}
 
LikeStarter is a social service that enables every registered user to spread his productions in the platform, e.g. songs, drawings, videos or just advertisements for a specific purpose. Published contents can receive the appreciation of users through the consolidated mechanisms typical of these platforms, i.e.~like, share and comment. Particular importance is given to the like feature, which, unlike other platforms, has a monetary as well as an involvement value. 
Thanks to this social form of appreciation, the artist will be able to earn funding and popularity, that will foster his future production. In this case, the tested crowdfunding mechanism used in Ethereum comes into play, in which every user ``like'' corresponds to a micro-donation.

\subsection{Like Donation}

Every time a donor likes a post, a certain amount of Ether is converted into tokens.
That is, one ``like'' in the system corresponds to a transaction that transfers a (small) amount of Ether to the crowdfunding beneficiary. In turn, the donor receives an amount of a token, introduced in LikeStarter, called Likoin.

In essence, the Likoin is the core of the crowdfunding mechanism, implemented as a standard ERC20 token~\cite{erc20}. The novel aspect of our approach is that the crowdsale is conveyed through the like mechanism; hence, its visibility is managed by the typical rules of social networks (e.g.~most liked posts are highly ranked).

\subsection{Crowdfunding}

LikeStarter employs a set of smart contracts to manage the crowdfunding without the need of a third entity. A user is the beneficiary of a crowdfunding and receives funds through likes to posts he produced or through his user page (same as a standard crowdfunding platform).

For a donor, possessing Likoins states that he has funded a particular user with a certain amount of Ether. Likoins can be employed to receive some benefits, i.e.~buying artifacts produced by the crowdfunding beneficiary (digital or physical products). Moreover, possessing Likoins enables the user with voting capabilities. Thus, the user gains an active role in the the crowdfunding management.

\subsection{Buying Artifacts}

To carry out a crowdfunding, a user can offer to his donors different products (here called artifacts). An artifact can only be traded for some Bucks, another kind of token introduced in the LikeStarter DAO, that can be acquired only by converting Likoin.
Hence, a crowdfunding beneficiary is bonded to two kinds of tokens:
\begin{itemize}
    \item Likoin: obtainable by a donor who funded the beneficiary (e.g. through a ``like'' to a post) with a certain amount of Ether. 
    \item Buck: obtainable by a donor who already owns some Likoins and decides to convert some of these Likoins to Bucks.
\end{itemize}
It is worth noticing that Likoins and Bucks refer to a single beneficiary, i.e.~Likoins and Bucks obtained for a beneficiary A, cannot be spent for the operations of a beneficiary B.

Based on these two tokens, the process of buying an artifact goes as follows. Once the artifact is produced and made available in the system, the donor can acquire some Likoins through a donation (which is paid in Ether). Then, he can convert some of his Likoins to Bucks. Finally, he trades some Bucks to acquire the artifact.

\subsection{Shares}
\label{share}

\begin{figure}[!h]
  \center
  \includegraphics[scale=0.135]{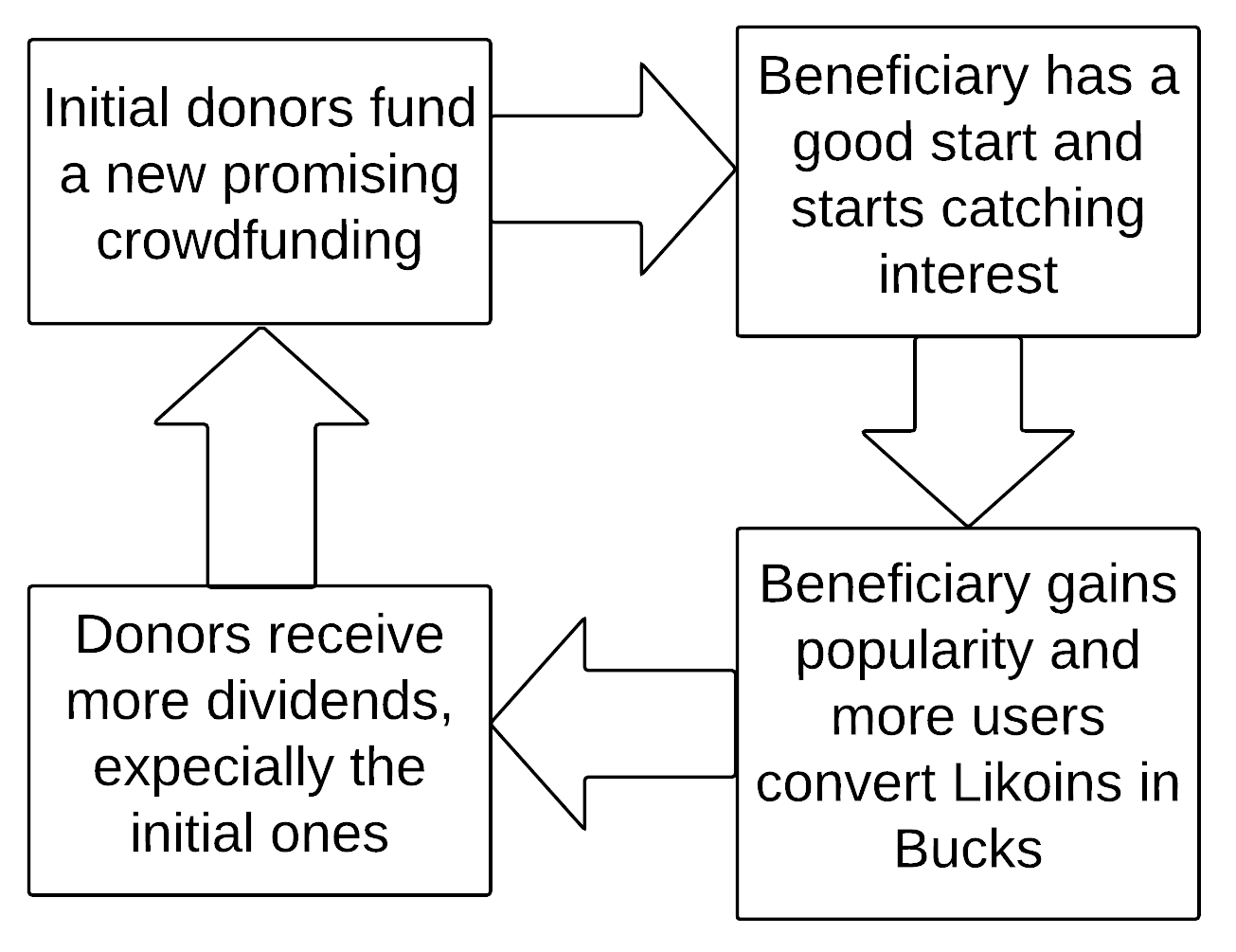}
  \caption{The autocatalytic cycle that incentivates the social behavior.}
  \label{fig:cycle}
\end{figure}

In LikeStarter, the price (in Bucks) of an artifact is not imposed by the crowdfunding beneficiary but is determined through a voting mechanism. For each artifact proposed by the beneficiary, every Likoin holder can suggest and/or vote a price. This means that the price is entirely chosen by voters. In this contest, an holder is driven to the choice of the right price by a fundamental rule: each Likoin that is converted in Buck is not burned, but it is distributed to all Likoin holders proportionally to their amount owned (e.g.~an user that holds 1\% of the existing Likoins for a specific beneficiary will receive 0.01 Likoin after 1 Likoin is converted to Buck by another donor). This approach permits to create an autocatalytic cycle where both donors and beneficiary increase their profits, as shown in Figure~\ref{fig:cycle}.

Thus, the Likoin can be seen as a derivative product, from which a donor can profit directly depending on the success of the financed artist~\cite{buterin2013}. This is achievable thanks to two factors: i) each conversion Likoin-Buck increments the amount of Likoins for an holder; and ii) a Likoin can be traded between two Ethereum accounts in exchange for Ether. In other words, the distribution and presence of Likoins make the LikeStarter application a DAO, in which Likoin holders, together with the beneficiary aim to the same objective: make the beneficiary ``famous'', increase the value of his artifacts and the value of its related Likoins.


\section{LikeStarter Software Architecture}\label{sec:archi}
 
\begin{figure}[!h]
  \center
  \includegraphics[scale=0.185]{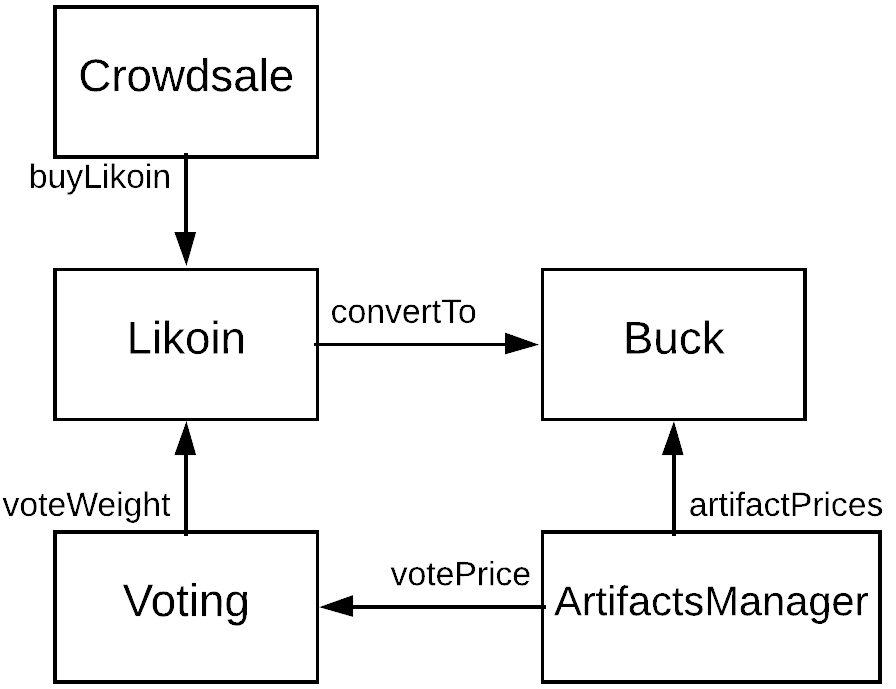}
  \caption{LikeStarter smart contracts architecture.}
  \label{fig:architecture}
\end{figure}

LikeStarter has been built as smart contracts based architecture, together with a social network application, that is also the Web interface needed to create a dialog between users and these contracts. The platform structure should be considered similar to a social network site (e.g.~Instagram) but that strongly relies on a set of smart contracts (see Figure~\ref{fig:architecture}).
 
 \subsection{Contracts}
 Contracts in LikeStarter allow to:
 \begin{itemize}
     \item Create and manage the two tokens and the crowdfunding;
     \item Manage the different artifacts;
     \item Orchestrate the DAO mechanisms as well as voting.
 \end{itemize}

\subsubsection{Likoin Smart Contract}
Likoin is the core component of the system, employed in almost every functionality of LikeStarter. It is implemented as a standard ERC20 token~\cite{erc20}. It acts as a token but it can be also seen as a financial share, expressing the ownership relationship between the crowdfunding beneficiary capital and the token holder. 
As a token, it can be transferred from an account to another or it can be minted (from the Crowdsale contract) to an account. It is also possible to delegate another account to manage a certain amount of tokens.
As a share, it represents how much a donor is influential to the beneficiary capital (i.e.~the ratio between the number of tokens owned by a donor and the total amount of minted tokens). More specifically, this influence is applied to the voting by means of the vote weight (that is equal to that ratio).

\subsubsection{Buck Smart Contract}
\begin{figure}[!h]
  \center
  \includegraphics[scale=0.24]{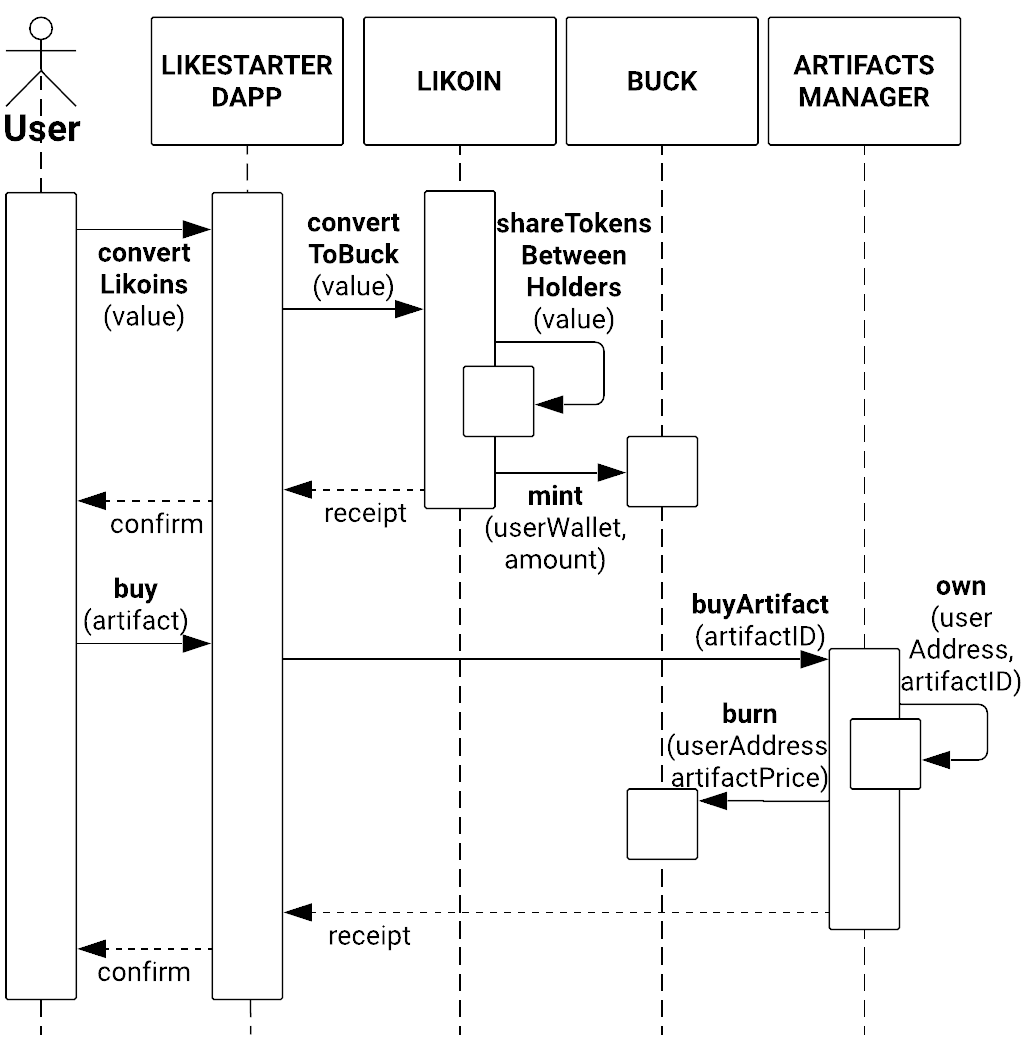}
  \caption{Sequence Diagram that illustrates how to convert Likoin in Bucks and to buy an artifact.}
  \label{fig:buyart}
\end{figure}

 Buck is not a standard ERC20 token, since it cannot be traded between accounts but it is only accepted for artifacts payments. This implies that once a Likoin has been converted to a Buck this operation is not reversible.  This token is necessary to let the Likoin be conceived as an ownership, rather than a currency strictly associated to the artifacts purchase. The duality of the two tokens allows the fundamental rule that enacts the cycle discussed in Section~\ref{share}. The interactions among users and smart contracts, during the Likoin conversion and artifact acquisition, is illustrated in Figure~\ref{fig:buyart}.
 
\subsubsection{Crowdsale Smart Contract}
\begin{figure}[!h]
  \center
  \includegraphics[scale=0.285]{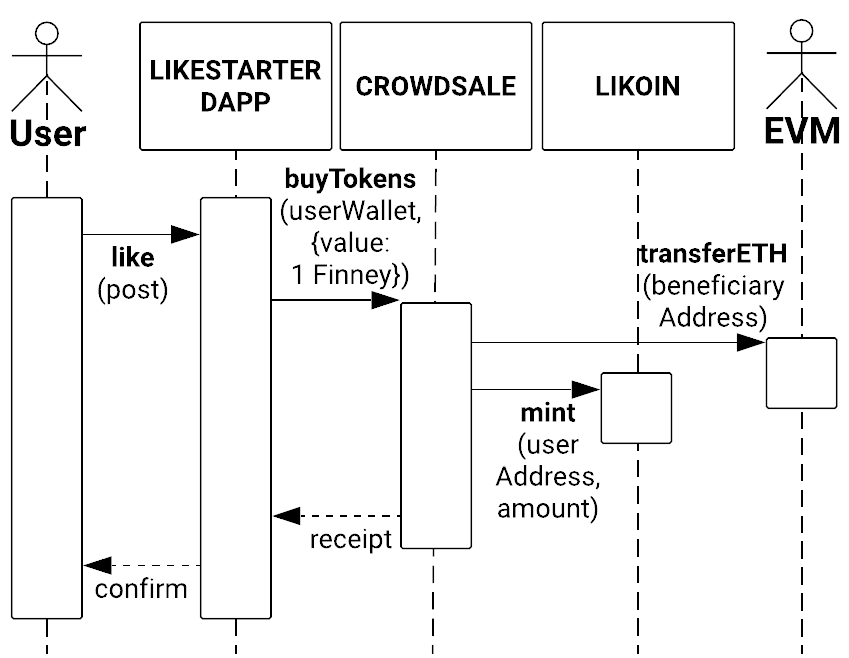}
  \caption{Sequence Diagram that illustrates how ``like'' works.}
  \label{fig:lik}
\end{figure}

 This is a simple contract, implementing the crowdfunding. It allows receiving Likoins, when transferring Ether to the beneficiary. It is thus used to mint Likoins. These Likoions are transferred to a donor, when he funds a beneficiary with a ``like'' in the social network application (that, in turn, corresponds to the transfer of a certain amount of Ether).
A sequence diagram of the ``like'' functioning is shown in Figure~\ref{fig:lik}.

\subsubsection{ArtifactsManager Smart Contract}
ArtifactsManager is a contract that allows the crowdfunding beneficiary to offer artifacts that can be traded for Bucks. This contract is used by the beneficiary to propose or remove some artifacts. This contract is used not only to maintain the beneficiary artifacts, but also to record the ownerships of an artifact to an account. There is a strong connection between this contract and the Voting contract because artifacts depend on the DAO.

\subsubsection{Voting Smart Contract}
\begin{figure}[!h]
  \center
  \includegraphics[scale=0.21]{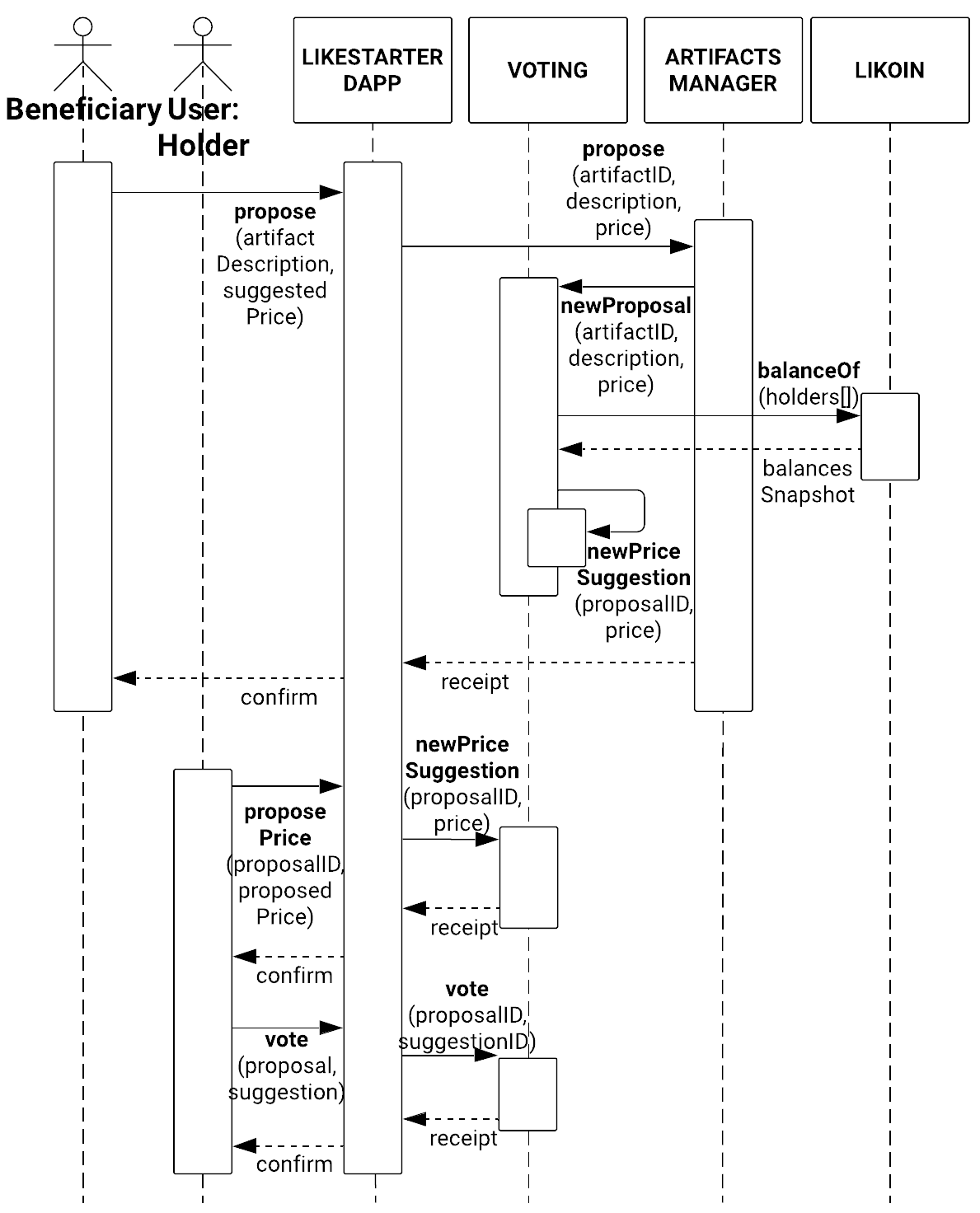}
  \caption{Sequence Diagram that illustrates how a beneficiary proposes an artifact and how a Likoin holder suggests a price and votes.}
  \label{fig:propose}
\end{figure}

The Voting contract can be seen as the DAO implementation. The voting system, in fact, regulates the behavior of the crowdfunding and of the related activities of the beneficiary. More specifically, this is the contract used to vote and reach a consensus on the value/price of an artifact. Only the crowdfunding beneficiary is allowed to propose an artifact through ArtifactsManager, but any other member can suggest and vote a price. A member is an account that holds any amount of Likoins (for that specific beneficiary) at the moment of the proposal. Each member vote is weighted, based on the stake that belongs to him. The interactions among users and smart contracts, during the voting process, is shown in Figure~\ref{fig:propose}.

\subsection{User interface}

This section describes the basic interactions between users and the LikeStarter social platform. The description is restricted to all basic functions and those involving the use of smart contracts. Every other function will be omitted.
Two roles for the user are considered: simple user and beneficiary user.
The most important concept is that for a simple user is possible, at any time, to open a crowdfunding campaign and become a beneficiary or, if that is not the case, the simple user can continue to use the platform, as in an usual social network site.

\subsubsection{Generic social network site interface}
The most important section of the site is the homepage, where most of the interactions between users are done. In essence, it is a list of posts that can be liked or commented and that include multimedia contents (e.g.~images, video, songs). Every user has a personal page containing his personal data, all the posts he has published or shared and a list of all donation that he has made in chronological order.

\subsubsection{Simple user} 
A simple user can use the following services:
\begin{itemize}
    \item Like a beneficiary post or offer some funds through the beneficiary personal social page. This is possible thanks to a transaction that calls the method to buy Likoins in the Crowdsale contract.
    \item Acquire a beneficiary artifact: an artifact can be acquired through Bucks, hence it is necessary to own some. To get Bucks, an user must generate a transaction that calls the method to convert Likoins in Bucks. 
    \item Suggest the price for an artifact through voting: every interaction in the DAO is managed throrugh transactions sent to the Voting contract.
    \item Start a crowdfunding campaign: the simple user will then become a beneficiary user and all the contracts will be deployed and assigned to him.
\end{itemize} 

A user can send transactions easily through a browser wallet (e.g.~MetaMask). 
Apart from this, the user interacts with the platform as in typical social networks.

\subsubsection{Beneficiary user}
In addiction to the services that a beneficiary inherits from the generic user, the following services are available:
\begin{itemize}
    \item Post content: a beneficiary user can support his crowdfunding through posts. Every post contributes to increase the beneficiary visibility, so that other users can appreciate his work and donate. Every donation is collected by the Crowdsale contract.
    \item Propose artifacts: every beneficiary can propose some artifacts in order to return something back to the ones that funded him. These artifacts are proposed to the DAO and their price is voted. This proposal is a simple transaction that calls a method provided by the ArtifactsManager.
    \item Suggest price and vote: a beneficiary is also a member of the DAO established through his contracts. Hence, he is able to suggest a price or vote as well as the others through the Voting contract.
\end{itemize}

A beneficiary can check the funds of his open campaign in his personal page. He does not need any transaction, all collected funds are displayed in Ether currency, with all information related to his campaign. Such information is  obtained from the blockchain. In this page, the beneficiary can also close his crowdfunding campaign, or check that status of votes for his artifacts.


\section{A Use Case with Artists Crowdfunding}\label{sec:use}

One of the main objectives we want to achieve with LikeStarter is to let anyone gain funds easily, reaching out as many people as possible. In this scenario, a typical use case could be an unknown artist that wants to emerge and receive funds.

In this section, we consider an application scenario in which a musician posts a song he produced. This enacts the voting process for the song's price within the DAO established around the musician.  Moreover, we also show how another user can make a donation to this musician. Thus, tasks performed are:
\begin{enumerate}
    \item Propose a new artifact;
    \item Suggest a price and vote for a suggestion;
    \item Make a donation.
\end{enumerate}

\begin{figure}[!h]
  \center
  \includegraphics[scale=0.160]{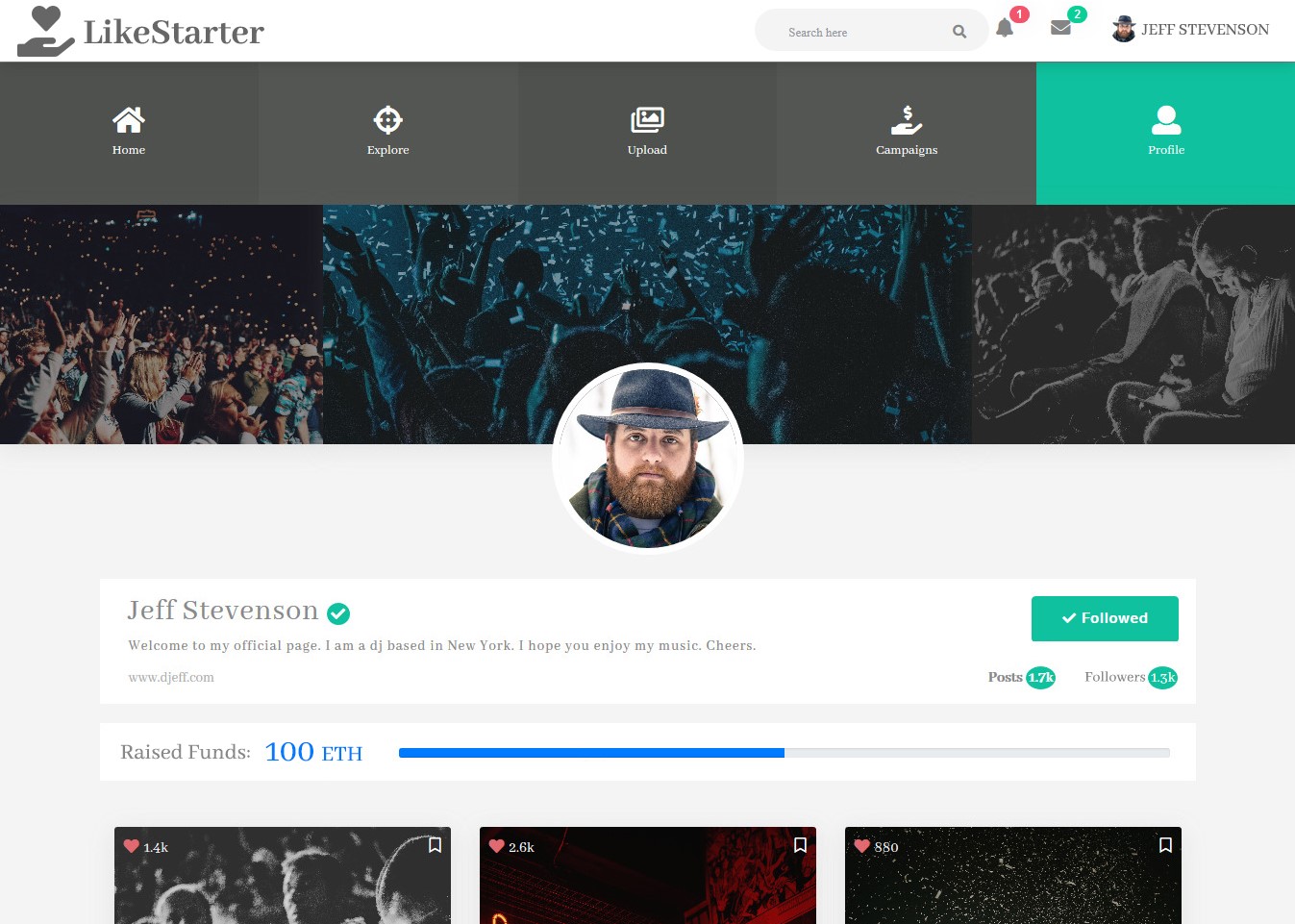}
  \caption{An artist public page in which is presented a recap of his personal informations, posts and donated funds.}
  \label{fig:artistpage}
\end{figure}

Let consider a musician, named Jeff Stevenson, already known in the platform. He started his crowdfunding by posting a song he made, catching the attention of different donors. Now, after he raised around 100 ETH (see Figure~\ref{fig:artistpage}) and published two albums through the platform, he wants to post a new single he produced for Christmas.

\subsection{Propose a new Artifact}
Through the Web page, Jeff decides to create a new proposal for an artifact, that his fans can buy with Bucks. Jeff gives a description of the song and suggests a price. Through his wallet application he confirms the transaction and the payment of fees. After the proposal submission (Figure~\ref{fig:propose}), the vote is open and every Likoin holder will receive a notification.

\subsection{Suggest a price and vote for a suggestion}
Every user, that owns Jeff's Likoins, is allowed to suggest (or vote for) a price. He can check all the details of Jeff's proposal. Then, he is able to submit a new price through a dedicated form. The transaction is submitted through his wallet application. Alternatively, he can vote for an existing suggestion. This operation requires also a transaction confirmation through the wallet application.

\subsection{Make a donation}
There are two ways to make a donation:
\begin{itemize}
    \item Click the ``like'' button of a post;
    \item Make a free donation in the artist personal page.
\end{itemize}

\begin{figure}[!h]
  \center
  \includegraphics[scale=0.160]{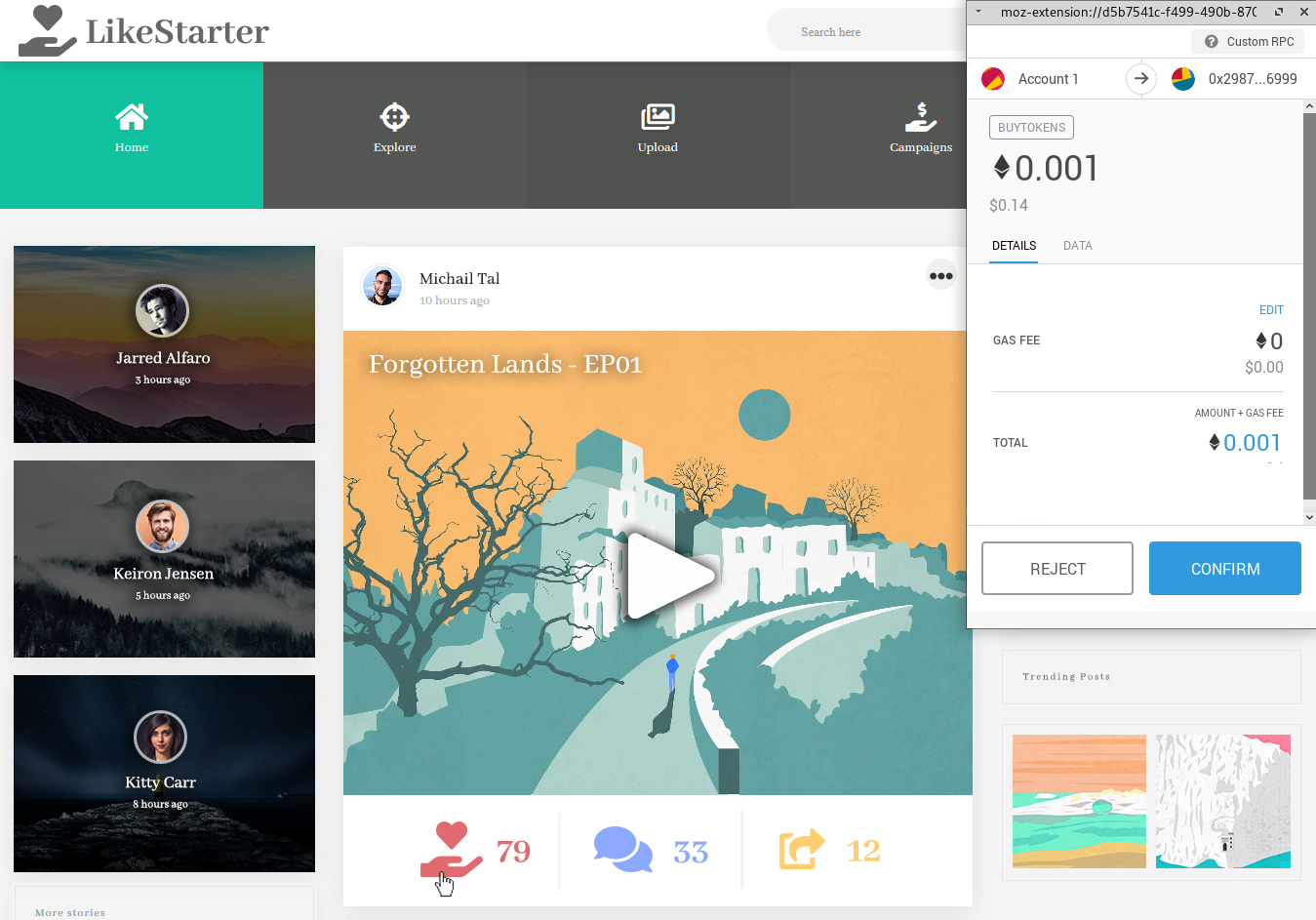}
  \caption{Screen in which is illustrated the donation through the like button.}
    \label{fig:likedonation}
\end{figure}

In both cases, a popup (created by the hot wallet) will appear on the browser to recap the transaction details (including fees), requiring a confirmation by the user (Figure~\ref{fig:likedonation}). This way, the ETH amount is transferred to the Crowdsale contract of the beneficiary (Jeff) and the donor user will get some Likoins in return. It is possible to check, at any moment, the amount of owned Likoins or the quantity of Bucks previously converted. 

\section{Conclusion}\label{sec:conc}

In this paper, we have discussed the main characteristics and features of LikeStarter, a novel social based DAPP for crowdfunding activities. The use of the Ethereum blockchain and smart contracts makes this application completely decentralized. 
LikeStarter assigns Likoins (i.e.~tokens related to an artist) to users that fund a given project. These tokens can be employed and converted to buy artifacts as well as token that provide users with voting capabilites. Users stake tokens on projects they support; in some sense, they possess a share on the beneficiary/project they are supporting. 
The more the value of this project/beneficiary the higher the value of these tokens.  
Users possessing Likoins gain also voting rights, i.e.~they can contribute in the decision of the price of certain artifacts.
All these aspects give to the application an autonomy that makes LikeStarter a DAO.

The source code of the LikeStarter project is available on GitHut
:~https://github.com/flamel13/eth-crowdsale.


\nocite{*}
\def\BibTeX{BibTeX}
\bibliographystyle{plain}
\bibliography{IEEEabrv,./biblio}

\begin{thebibliography}{10}

\bibitem{NBERw19133}
Ajay~K Agrawal, Christian Catalini, and Avi Goldfarb.
\newblock Some simple economics of crowdfunding.
\newblock Working Paper 19133, National Bureau of Economic Research, June 2013.

\bibitem{buterin2013}
Vitalik Buterin.
\newblock Ethereum: A next-generation smart contract and decentralized
  application platform.
\newblock 2013.

\bibitem{cai2018decentralized}
Wei Cai, Zehua Wang, Jason~B Ernst, Zhen Hong, Chen Feng, and Victor~CM Leung.
\newblock Decentralized applications: The blockchain-empowered software system.
\newblock {\em IEEE Access}, 6:53019--53033, 2018.

\bibitem{gda-jpdc-2017}
Gabriele D'Angelo and Stefano Ferretti.
\newblock Highly intensive data dissemination in complex networks.
\newblock {\em Journal of Parallel and Distributed Computing}, 99:28 -- 50,
  2017.

\bibitem{D'Angelo:2018}
Gabriele D'Angelo, Stefano Ferretti, and Moreno Marzolla.
\newblock A blockchain-based flight data recorder for cloud accountability.
\newblock In {\em Proc. of the 1st Workshop on Cryptocurrencies and Blockchains
  for Distributed Systems}, CryBlock'18, New York, NY, USA, 2018. ACM.

\bibitem{8327568}
G.~Fenu, L.~Marchesi, M.~Marchesi, and R.~Tonelli.
\newblock The ico phenomenon and its relationships with ethereum smart contract
  environment.
\newblock pages 26--32, March 2018.

\bibitem{gerber}
Elizabeth Gerber, Julie Hui, and Pei-Yi~(Patricia Kuo.
\newblock Crowdfunding: Why people are motivated to post and fund projects on
  crowdfunding platforms.
\newblock {\em Computer Supported Cooperative Work 2012, Workshop on Design
  Influence and Social Technologies: Techniques, Impacts and Ethics, Seattle,
  WA.}, 10, 02 2012.

\bibitem{GLEASURE2016101}
Rob Gleasure and Joseph Feller.
\newblock Emerging technologies and the democratisation of financial services:
  A metatriangulation of crowdfunding research.
\newblock {\em Information and Organization}, 26(4):101 -- 115, 2016.

\bibitem{jentzsch2016}
Christoph Jentzsch.
\newblock Decentralized autonomous organization to automate governance.
\newblock 2016.

\bibitem{KIM2018153}
Ji~Won Kim.
\newblock They liked and shared: Effects of social media virality metrics on
  perceptions of message influence and behavioral intentions.
\newblock {\em Computers in Human Behavior}, 84:153 -- 161, 2018.

\bibitem{chunta}
Chun-Ta Lu, Sihong Xie, Xiangnan Kong, and Philip~S. Yu.
\newblock Inferring the impacts of social media on crowdfunding.
\newblock pages 573--582, 2014.

\bibitem{Seijas2016ScriptingSC}
Pablo~Lamela Seijas, Simon~J. Thompson, and Darryl McAdams.
\newblock Scripting smart contracts for distributed ledger technology.
\newblock {\em IACR Cryptology ePrint Archive}, 2016:1156, 2016.

\bibitem{erc20}
Fabian Vogelsteller and Vitalik Buterin.
\newblock Eip 20: Erc-20 token standard.

\bibitem{article}
Aaron Wright and Primavera De~Filippi.
\newblock Decentralized blockchain technology and the rise of lex
  cryptographia.
\newblock {\em SSRN Electronic Journal}, 01 2015.

\end{thebibliography}

\end{document}